\documentclass[physrev]{revtex4-2}

\usepackage{amsthm, amssymb}
\theoremstyle{definition}
\newtheorem{definition}{Definition}
\newtheorem{theorem}{Theorem}

\usepackage{graphicx}
\usepackage{dcolumn}
\usepackage{bm}
\usepackage{centernot}
\usepackage{amsmath}

\usepackage{enumitem}
\usepackage[normalem]{ulem}
\usepackage{url}

\usepackage{hyperref}
\hypersetup{colorlinks,breaklinks,
			citecolor=[rgb]{0,0.0,1.0},
            urlcolor=[rgb]{0.0,0.0,1.0},
            linkcolor=[rgb]{0,0.5,0.9}}

\begin{document}

\title{\large{Covariant conservation laws, local invariance and Noether’s second theorem}} 

\author{Nuno Barros e S\'{a}}
\email{nuno.bf.sa@uac.pt}
\affiliation{\small Faculdade de Ci\^encias e Tecnologia, Universidade dos A\c{c}ores, 9500-321 Ponta Delgada, Portugal,}
\affiliation{\small Instituto de Astrof\'{i}sica e Ci\^{e}ncias do Espa\c{c}o, Faculdade de Ci\^encias, Universidade de Lisboa, Campo Grande,  Edif\'{i}cio C8, 1749-016 Lisboa, Portugal.}

\author{Miguel A. S. Pinto}
\email{mapinto@fc.ul.pt}
\affiliation{\small Instituto de Astrof\'{i}sica e Ci\^{e}ncias do Espa\c{c}o, Faculdade de Ci\^encias, Universidade de Lisboa, Campo Grande,  Edif\'{i}cio C8, 1749-016 Lisboa, Portugal,}
\affiliation{\small Departamento de F\'{i}sica, Faculdade de Ci\^{e}ncias da Universidade de Lisboa,  Edif\'{i}cio C8, Campo Grande, P-1749-016 Lisbon, Portugal,}
\affiliation{\small William H. Miller III Department of Physics and Astronomy, Johns Hopkins University, 3400 North Charles Street, Baltimore, Maryland, 21218, USA.}

\author{Tom\'as Trindade}
\email{ttrindadehenriques@gmail.com}
\affiliation{\small Centro de F\'{\i}sica Te\'{o}rica e Computacional, Faculdade de Ci\^encias, Universidade de Lisboa, Campo Grande,  Edif\'{i}cio C8, 1749-016 Lisboa, Portugal,}
\affiliation{\small Departamento de F\'{i}sica, Faculdade de Ci\^{e}ncias da Universidade de Lisboa,  Edif\'{i}cio C8, Campo Grande, P-1749-016 Lisbon, Portugal.\\}

\date{\today}

\begin{abstract}
We lay down a set of requirements for a field theory to produce a covariant conservation law out of Noether’s second theorem, and show that neither local invariance implies a covariant conservation law, nor the existence of a covariant conservation law necessarily stems from the local invariance of the theory. We illustrate our results with the examples of well-known theories.
\end{abstract}


\maketitle
\thispagestyle{empty}


\setcounter{page}{1} 

\section{Introduction}\label{intro}

Covariant conservation laws may arise in field theories where a notion of parallel transport, that is, a connection, can be defined. We begin by defining what is meant by conservation laws. Throughout the article, the symbol $\approx$ means equality when the equations of motion are satisfied (on-shell) and the symbol $=$ means absolute equality (off-shell).

\begin{definition} A proper conservation law is a relation of the type $\partial_\mu j^\mu\approx0$, with $j^\mu$ being a local function of the fundamental fields of the theory. For the law not to be trivial, $j^\mu$ must not be identically zero on-shell and $\partial_\mu j^\mu$ must not be identically zero off-shell.
\end{definition}

\begin{definition} A covariant conservation law is a relation of the type $\nabla_\mu j^\mu_a\approx0$, where $j^\mu_a$ is a local section of $T{\cal M}\otimes E$, with $T{\cal M}$ the tangent bundle of the manifold $\cal M$, $E$ an associated vector bundle of the theory, $\nabla_\mu$ the covariant derivative induced by the relevant connection on $E$, and the index $a$ is a fiber index of $E$. For the covariant conservation law not to be trivial, $j^\mu_a$ must not be identically zero on-shell and $\nabla_\mu j^\mu_a$ must not be identically zero off-shell.
\end{definition}

Covariant conservation laws show up in the most widely used field theories. In General Relativity, the energy-momentum tensor of matter obeys the covariant conservation equation
\begin{equation}
    \nabla_{\mu}T^{\mu \nu}=\partial_{\mu}T^{\mu \nu}+\Gamma_{\mu\alpha}{^\mu}T^{\alpha\nu}+\Gamma_{\mu\alpha}{^\nu}T^{\mu\alpha}\approx 0 \text{ ,}\label{gr2}
\end{equation}
where $\nabla_{\mu}$ is the covariant derivative with respect to the Levi-Civita connection $\Gamma_{\mu\alpha}{^\nu}$. Similarly, in Yang-Mills theories, the current of matter obeys the equation of covariant conservation
\begin{equation}
\nabla_{\mu}j_{a}^{\mu}=\partial_{\mu}j_{a}^{\mu}+ g_{\text{YM}} C_{abc}A_\mu^bj_c^{\mu}\approx 0 \text{ ,}\label{ym2}
\end{equation}
where $\nabla_{\mu}$, $g_{\text{YM}}$, $C_{abc}$, and $A_{\mu }^{a}$ are the covariant derivative (here, in the adjoint representation), coupling constant, structure constants, and vector potential of the gauge group in question.

In both cases, the theories are invariant under groups of local transformations, and Noether’s second theorem can be used to obtain the covariant conservation laws. The recent interest in alternative theories, particularly to General Relativity, raises the question of whether such a relation between local invariances and covariant conservation laws is general.

Noether's two theorems \cite{noe} concern integrals of local functions that are invariant under groups of transformations. Noether's first theorem states that ``for each {\sl global} parameter of the group involved in the invariance of an integral, there is a current whose divergence is a combination of the Lagrange expressions resulting from this integral''. Noether's second theorem states that ``for each {\sl local} parameter of the group involved in the invariance of an integral, there exists a relation between its Lagrange expressions (and possibly their derivatives)''.

When the integral in both Noether's theorems is the action for a physical theory, the equations of motion are obtained by setting the Lagrange expressions to zero. In this case, Noether's first theorem provides equations of continuity, such as the conservation of the energy-momentum tensor $T^{\mu\nu}$ or the conservation of charge, respectively
\begin{equation}
    \partial_{\mu}T^{\mu \nu}\approx 0 \quad\text{and}\quad\partial_{\mu}j_{a}^{\mu}\approx 0 \text{ .}
    \label{gr}
\end{equation}
This is her most famous result, and in many physicists' minds, this is what Noether's theorem is about. The similarity between Eqs. \eqref{gr2} and \eqref{ym2} on one hand, and Eqs. \eqref{gr} on the other hand, may make one believe that the former equations arise as a consequence of Noether's first theorem too. However, that cannot be the case, as Noether's first theorem concerns global symmetries, not local symmetries (as is the case in General Relativity and Yang-Mills theories).

\medskip

\noindent{\bf Remark} {\it Noether's first theorem provides laws of pro\-per conservation, not laws of covariant conservation.}

\medskip

On the other hand, Noether's second theorem, valid for local symmetries, when applied to the action of a physical theory, and when the symmetry transformations affect only the variable fields of the theory and possibly the underlying spacetime, provides off-shell relations between the equations of motion, meaning that these are not independent, and therefore that it is not possible to completely solve them to obtain the fields, which is a manifestation of the gauge freedom associated with the invariance under a group of local transformations. This is the most renowned consequence of Noether's second theorem.

\medskip

\noindent{\bf Remark} {\it Noether's second theorem, when applied to the invariance of the action of a theory under transformations of its spacetime and of its variable fields, provides relations between the equations of motion, not laws of covariant conservation.}

\medskip

But, as we shall see, when applied under different conditions, Noether's second theorem can be used to generate laws of covariant conservation \cite{nun}.

This is often a source of confusion in the literature. For example, it is sometimes stated that the covariant conservation of the energy-momentum of matter in General Relativity is a direct consequence of the diffeomorphism invariance of the theory \cite{carr,car}. Some authors acknowledge that the extra ingredient of separability of the action of the theory into two parcels, one containing the matter fields and the other not, present in the theory of General Relativity, is necessary to derive the law of covariant conservation of the energy-momentum tensor of matter from Noether's second theorem \cite{haro,hob,for}. Here we set up the general situation, with its application to modern alternative theories in sight.

Our main results are:

\begin{enumerate}[label=\arabic*.]
\item Laws of covariant conservation can be derived from Noether's second theorem for gauge theories and for metric diffeomorphism invariant theories if one of the three following conditions is satisfied by the ``radiation'' fields $\zeta_I$ (the gauge potentials or the metric): 
\begin{enumerate}[label=\roman*)]
    \item The invariant integral is the whole action of the theory, but the fields $\zeta_I$ are background fields, i.e., prescribed functions of spacetime, not variables.
    \item The invariant integral is one of two terms of the action, the other term involving only the fields $\zeta_I$.
    \item The invariant integral is the action, but its integrand contains combinations of the fields $\zeta_I$ which are local invariant themselves.
\end{enumerate}
When one of these conditions is met for fields $\zeta_I$, but these fields are not the gauge potentials or the metric, non-trivial on-shell identities can still be derived from Noether's second theorem, but they do not express covariant conservation laws.
\item As a consequence, neither does local invariance of a theory imply the existence of a covariantly conserved current, nor does the existence of a covariantly conserved current denounce an underlying local symmetry of the theory.
\end{enumerate}

The paper is organized as follows. In Sec. \ref{sec2}, we review Noether's second theorem when applied to the action of the theory. In Sec. \ref{sec3}, we establish the conditions under which Noether's second theorem can be used to produce relevant identities valid on-shell. In Sec. \ref{sec4}, we show that for gauge theories and for metric diffeomorphism invariant theories, these identities can be expressed using covariant derivatives. Finally, in Sec. \ref{sec5}, we establish the conditions for laws of covariant conservation to emerge in these theories: they fall into three categories. In sec. \ref{sec6} we give examples of theories falling into each one of these categories. They include Yang-Mills theory and General Relativity, and more general gauge and diffeomorphism invariant metric theories, with the gauge connection or the metric being either external fields or variable fields but with a kinetic term in the Lagrangian. We also include a toy model. In Sec. \ref{sec7}, we give examples of theories with local invariance where no law of covariant conservation emerges from Noether's second theorem: metric-affine theories, Poincar\'e gauge theory, Weyl conformal gravity and $f(R, T)$ gravity. In Sec. \ref{sec8}, we discuss alternative approaches to arrive at covariantly conserved currents apart from Noether's second theorem.

\section{Euler-Lagrange expressions and Noether's second theorem}
\label{sec2}

Given the integral $\cal S$ of a local functional $\cal L$ of the fields $\phi _{I}$, with $I=1,\dots,N$, defined on a $D$-dimensional subset $\mathcal{V}$ of a smooth manifold $\mathcal{M}$ (with ``local'' meaning that it may depend on these $N$ fields and their derivatives up to a finite order $n$),
\begin{equation}
{\cal S}=\int_{\mathcal{V}} {\cal L}\left( \phi _{I}\right) \, d^Dx \text{ ,}\label{int1}
\end{equation}%
an infinitesimal variation of the fields $\delta \phi _{I}$ produces the variation of the integral
\begin{equation}
\delta {\cal S}=\int_{\mathcal{V}}\frac{\delta {\cal L}}{\delta \phi _{I}}\delta \phi _{I} \, d^Dx
\label{bum}
\end{equation}
plus boundary terms (which we omitted since they are not going to be relevant for the discussion), with the Euler-Lagrange derivatives given by
\begin{equation}
\frac{\delta {\cal L}}{\delta \phi _{I}}=\sum_{i=0}^{n}\left( -1\right) ^{i}\partial _{\mu _{i}}\ldots \partial
_{\mu _{1}}\frac{\partial {\cal L}}{\partial \left( \partial _{\mu _{i}}\cdots
\partial _{\mu _{1}}\phi _{I}\right)}=\frac{\partial {\cal L}}{\partial \phi _{I}}-\partial _\mu\frac{\partial {\cal L}}{\partial \left(\partial _\mu\phi _{I}\right)}+\partial _\mu\partial _\nu\frac{\partial {\cal L}}{\partial \left(\partial _\mu\partial _\nu\phi _{I}\right)}-\cdots\text{ .}\label{pupu}
\end{equation}
Here, and in the remaining of the article, it is understood that indices must be numbered from right to left, that is, $\partial_{\mu_3}\dots\partial_{\mu_1}\phi=\partial_{\mu_3}\partial_{\mu_2}\partial_{\mu_1}\phi$, $\partial_{\mu_1}\dots\partial_{\mu_1}\phi=\partial_{\mu_1}\phi$, but $\partial_{\mu_0}\dots\partial_{\mu_1}\phi=\phi$.

Now, let us consider a group of local transformations parameterized by a finite number of functions $\varepsilon^{a}$ ($a=1,\dots,A$, with $A$ being the dimension of the continuous symmetry group) such that an infinitesimal transformation of the fields depends on $\varepsilon^{a}$ and their derivatives up to some finite order $m$,
\begin{equation}
\delta \phi _{I} =\sum_{i=0}^{m}f_{Ia}^{\mu _{i}\ldots \mu _{1}}\partial _{\mu _{i}}\cdots
\partial _{\mu _{1}}\varepsilon^{a}\text{ ,}\label{transf}
\end{equation}
where the $f_{Ia}^{\mu _{i}\ldots \mu _{1}}$ are some functions, possibly
depending on the fields $\phi _{I}$ and their derivatives. Since the $\varepsilon^{a}$ are arbitrary functions, one can choose them such that their derivatives up to a
sufficiently high finite order vanish at the boundary of the domain of integration $\partial \mathcal{V}$ (that being the reason why we ignored boundary terms in Eq. \eqref{bum}), which then becomes
\begin{equation}
\delta {\cal S}=\int_{\mathcal{V}}\frac{\delta {\cal L}}{\delta \phi _{I}}\sum_{i=0}^{m}f_{Ia}^{\mu
_{i}\ldots \mu _{1}}\partial _{\mu _{i}}\cdots \partial _{\mu
_{1}}\varepsilon^{a} \, d^Dx \text{ .}\label{eq9}
\end{equation}%
Further integration by parts can be performed, yielding, ignoring the resulting boundary terms,
\begin{equation}
\delta{ {\cal S}}=\int_{\mathcal{V}}\sum_{i=0}^{m}\left( -1\right) ^{i}\partial _{\mu
_{i}}\cdots \partial _{\mu _{1}}
\left( \frac{\delta {\cal L}}{\delta \phi _{I}}%
f_{Ia}^{\mu _{i}\ldots \mu _{1}}\right)\varepsilon^{a} \, d^Dx \text{ .}\label{eq10}
\end{equation}
In addition, if we assume that the integral is invariant under this group of transformations (or, more generally, that its variation produces at most a boundary term), ${\cal S}=S_{\rm inv}=\int_{\cal V}L_{\rm inv}\,d^Dx$, then $\delta S_{\rm inv}=0$ implies%
\begin{equation}
\sum_{i=0}^{m}\left( -1\right) ^{i}\partial _{\mu _{i}}\cdots \partial _{\mu
_{1}}\left( \frac{\delta L_{\rm inv}}{\delta \phi _{I}}f_{Ia}^{\mu _{i}\ldots \mu
_{1}}\right) =0\label{bquatro}
\end{equation}%
for $a=1,\dots,A$. These are $A$ relations between the Lagrange expressions, which constitute the content of Noether's second theorem.

When the integral ${\cal S}$ is the action $S$ of a physical theory, that is, ${\cal S}=S=\int_{\cal V} L\ d^Dx$, $L$ being the corresponding Lagrangian density, and the fields $\phi_I$ are the variable fields of the theory, the equations of motion are obtained from the principle of stationary action, demanding in Eq. \eqref{bum} that $\delta \phi_I$ vanish at the boundary of the domain of integration for all fields (together with its derivatives up to order $n-1$)
\begin{equation}
\delta S=0\quad \Rightarrow \quad \frac{\delta L}{\delta \phi _{I}}\approx 0 \text{ .}\label{emo}
\end{equation}
In other words, the equations of motion are given by setting the Lagrange expressions to zero, and Eqs. \eqref{bquatro} that arise from Noether's second theorem are a set of $A$ relations between the equations of motion.

As an example, below we provide a simple application of Noether's second theorem to the action of a Yang-Mills theory.

\subsection{An illustrative example: Yang-Mills theory}
A Yang-Mills theory with a vector potential $A_{\mu }^{a}$ coupled to a
Dirac field $\psi $ is described by the 4-dimensional action
\begin{equation}
S =\int_{\mathcal{V}} \left[ -\frac{1}{4}F_{\mu \nu }^{a}F_{a}^{\mu \nu }+\frac{\rm i}{2}\left( 
\bar{\psi}\gamma ^{\mu }\nabla_{\mu }\psi- \overline{\nabla_{\mu }\psi} \gamma ^{\mu }\psi\right)-m\bar{\psi}\psi \right] \, d^4x\text{ ,}
\end{equation}
where the covariant derivative of the Dirac field and the strength field tensor $F_{\mu \nu }^{a}$ are defined as
\begin{eqnarray}
\nabla_{\mu }\psi&=& \partial _{\mu }\psi
-{\rm i}g_{\text{YM}}A_{\mu }^{a}T_{a}\psi \text{ ,}\\
F_{\mu \nu }^{a}&=&\partial _{\mu }A_{\nu }^{a}-\partial _{\nu }A_{\mu
}^{a}+g_{\text{YM}}C_{abc}A_{\mu }^{b}A_{\nu }^{c}\text{ ,}
\end{eqnarray}%
with $g_{\text{YM}}$, $C_{abc}$ and $T_a$ being, respectively, the coupling constant, the structure constants and the generators of the gauge group defined by the algebra $\left[ T_{a},T_{b}\right] ={\rm i}C_{abc}T_{c}$.

Accordingly, by varying the action with respect to $\bar{\psi}$ and $A_{\mu}^{a}$, we obtain the equations of motion for $\psi$ and $A_{\mu}^{a}$, respectively
\begin{eqnarray}
\frac{\delta L}{\delta \bar{\psi}}&=&{\rm i}\gamma ^{\mu }\nabla_{\mu }\psi
-m\psi \approx 0 \text{ ,}\label{si}\\
\frac{\delta L}{\delta A_\mu^a}&=&\nabla _{\nu }F_{a}^{\nu \mu
}+g_{\text{YM}}\bar{\psi}\gamma ^{\mu
}T_{a}\psi\approx 0\text{ ,}\label{si2}
\end{eqnarray}%
and we note that the equation of motion for adjoint field $\bar\psi$ is the adjoint of Eq. \eqref{si}. Furthermore, this theory is invariant under gauge transformations, which acquire the infinitesimal form
\begin{eqnarray}
\delta\psi&=&{\rm i}g_{\text{YM}}\varepsilon^aT_{a}\psi\label{ymg} \text{ ,}\\
\delta A^{a}_{\mu}&=&\partial _{\mu }\varepsilon
^{a} + g_{\text{YM}} C_{abc}A_{\mu}^b\varepsilon^{c}=\nabla_{\mu }\varepsilon^{a}\text{ .}  \label{cd2}
\end{eqnarray}%
Thus, Noether's second
theorem applied to the whole action provides us with the following relations associated with this internal gauge
symmetry%
\begin{equation}
{\rm i}g_{\text{YM}}\left(\frac{\delta L}{\delta \psi } T_a \psi -\bar{\psi} T_a \frac{\delta L}{\delta \bar{%
\psi}}\right)-\nabla_{\mu }\frac{\delta L}{\delta A_{\mu }^{a}}=0\text{ ,}\label{snym}
\end{equation}
which can be easily verified to hold by plugging in Eqs. \eqref{si}-\eqref{si2}.

\section{Non-trivial on-shell identities from Noether's second theorem}\label{sec3}
Noether's second theorem can be applied to any invariant integral, not necessarily the action of a physical theory, and even if the group of transformations acts on background fields besides the variables of the theory, in which cases the Lagrange expressions do not necessarily vanish on-shell. Then, Noether's second theorem provides non-trivial identities for the theory which are valid on-shell. 

\begin{theorem}\label{teo1}
Let
$S_{\rm inv} = \int_{\mathcal{V}} L_{\rm inv}\left( \phi _{I}\right) \, d^Dx$
be an integral that is invariant under an $A$-dimensional group of local transformations of the fields $\phi_{I}$. If it is possible to separate the fields $\phi_{I}$ into two subsets, $\zeta_I$ ($I=1,...,M$, with $0<M<N$) and $\psi_I$ ($I=1,...,N-M$), in such a manner that the action $S= \int_{\mathcal{V}} L\left( \phi _{I}\right) \, d^Dx$ of a physical theory is related to $S_{\rm inv}$ according to one of the conditions 1–3 listed below, then Noether's second theorem~\eqref{bquatro} produces $A$ non-trivial on-shell identities $\Psi_a \approx 0$ ($a=1,\ldots ,A$), with $\Psi_a$ not being identically zero off-shell. The conditions are:

\begin{enumerate}[label=\arabic*.]
    \item The action of the theory is the invariant integral, $S=S_{\rm inv}[\zeta_I,\psi_I]$, with the fields $\psi_I$ being variables of the theory and the fields $\zeta_I$ being background fields, i.e., prescribed functions of spacetime, not variables.
    \item All the fields are variables, but the action of the theory is the sum of two terms, the invariant integral plus another term involving only the fields $\zeta_I$, $S=S_{\rm inv}[\zeta_I,\psi_I]+S_{\rm g}[\zeta_I]$.
    \item All the fields are variables, the action of the theory is the invariant integral, and its integrand contains combinations ${\cal I}_p$ of the fields $\zeta_I$ which are local invariant themselves, $S=S_{\rm inv}[\psi_I,\zeta_I,{\cal I}_p(\zeta_I)]$.
\end{enumerate}
\end{theorem}

\begin{proof}
For the three cases:
    \begin{enumerate}[label=\arabic*.]
    \item From Eq. \eqref{bquatro}, one has
    \begin{equation}
\sum_{i=0}^{m}\left( -1\right) ^{i}\partial _{\mu _{i}}\cdots \partial _{\mu
_{1}}\left( \frac{\delta L_{\rm inv}}{\delta \zeta_{I}}f_{Ia}^{\mu _{i}\ldots \mu
_{1}}\right)+\sum_{i=0}^{m}\left(-1\right) ^{i}\partial _{\mu _{i}}\cdots \partial _{\mu
_{1}}\left(\frac{\delta L_{\rm inv}}{\delta \psi _{I}}f_{Ia}^{\mu _{i}\ldots \mu
_{1}}\right)=0\text{ .}
\end{equation}
Since the fields $\psi_I$ are variables of the theory, $\delta L_{\rm inv}/\delta\psi_I\approx0$ holds. Then
    \begin{equation}
\Psi_a=\sum_{i=0}^{m}\left( -1\right) ^{i}\partial _{\mu _{i}}\cdots \partial _{\mu
_{1}}\left( \frac{\delta L_{\rm inv}}{\delta \zeta_{I}}f_{Ia}^{\mu _{i}\ldots \mu
_{1}}\right)\approx 0\text{ .}\label{bquatro2}
\end{equation}
But $\Psi_a$ is not identically zero off-shell because $\delta L_{\rm inv}/\delta \psi_{I}$ is not. 
However, the background fields $\zeta_I$ do not produce equations of motion, hence, in the general case, $\delta L_{\rm inv}/\delta \zeta_{I}$ is not identically zero on-shell, and the identity $\Psi_a\approx0$ is not trivial. 
    \item With $S_{\rm g}[\zeta_I]=\int L_{\rm g}(\zeta_I)d^Dx$, the equations of motion for the fields $\psi_I$ are $\delta L_{\rm inv}/\delta \psi_{I}\approx 0$, but for the fields $\zeta_I$ they are $\delta (L_{\rm inv}+L_{\rm g})/\delta \zeta_{I}\approx 0$. Therefore, in the general case, $\delta L_{\rm inv}/\delta \zeta_{I}\approx-\delta L_{\rm g}/\delta \zeta_{I}$ is not identically zero on-shell. Then the proof follows the same steps as in the previous case.
    \item One has
\begin{equation}
\frac{\delta L_{\rm inv}}{\delta \zeta _{I}}\delta \zeta _{I}=\frac{\delta L_{\rm inv}}{\delta {\cal I}_{p}}\delta {\cal I}_{p}+\left(\frac{\delta L_{\rm inv}}{\delta \zeta _{I}}\right)_{{\cal I}_p={\rm const}}\delta \zeta _{I}
\end{equation}
Because $\delta {\cal I}_p=0$, the first term on the right-hand side of this equation is zero, and only the term
\begin{equation}
    \left(\frac{\delta L_{\rm inv}}{\delta \zeta_{I}}\right)_{{\cal I}_p={\rm const}}{\text ,}
\end{equation}
which is not identically zero on-shell, shows up in Noether's second theorem, Eq. \eqref{bquatro}. Then, the proof follows the same steps.
\end{enumerate}
\end{proof}

\noindent{\bf Remark} {\it In Theorem \ref{teo1} the integrand of the invariant integral either plays the role of the total Lagrangian of the theory (cases 1 and 3), $L_{\rm inv}=L$, or of one of its terms, usually its matter part, $L_{\rm inv}=L_{\rm m}$ (case 2). The fields $\psi_I$ and $\zeta_I$ are typically associated with the ``matter'' and ``radiation'' fields, respectively.}

\medskip

In Section \ref{sec6}, the reader may find explicit examples of the three cases mentioned in Theorem \ref{teo1}.

\section{Gauge theories and metric diffeomorphism invariant theories}\label{sec4}

The identities of the previous section are not, for general local transformations, laws of covariant conservation. For that to happen, it is necessary that the fields in the theory form fibers $E$ over each point $x\in{\cal M}$ of the manifold, and that fibers are not canonically identifiable across points. Then, the theory must specify a connection that provides for parallel transport (and hence a covariant derivative).

That happens with gauge transformations, the connection being either a variable field or a background field. Then one can take the parameters $\varepsilon^a$ of the gauge transformation in Eq. \eqref{transf} to be the coordinates (varying from point to point in spacetime) of an element of the Lie algebra in the adjoint representation of the group that leaves the integral Eq. \eqref{int1} invariant.

We shall also consider diffeomorphisms in spacetime when a metric (or a tetrad field) is present, either as a variable or as a background field. Then the parameters $\varepsilon^\mu$ are coordinates of the Lie algebra-valued generators of diffeomorphisms (vector fields) in the basis $\partial_\mu$. The spacetime connection may be one of the fields of the theory, but, even in its absence, the metric guarantees the existence of a natural connection, the Levi-Civita connection, on which one can build a covariant derivative operator. We shall not consider diffeomorphism invariant theories without a metric structure, as such theories are typically purely topological, with no propagating degrees of freedom.

\medskip

\noindent{\bf Remark} {\it In the following, we shall treat both cases (gau\-ge theories and metric theories) with the same terminology: $\varepsilon^a$ for the gauge parameters and $\nabla_\mu$ for the covariant derivative, with the understanding that, in the case of the group of diffeomorphisms, the index $a$ stands for a spacetime index, and that, in the case of a gauge group, the symbol $\nabla$ stands for the covariant derivative with respect to the gauge connection.}

\begin{theorem}\label{teo2}
When the group of invariance is an internal gauge group or the group of diffeomorphisms in a metric theory, the following results hold:
\begin{enumerate}[label=\arabic*.]
    \item The infinitesimal transformation of the fields Eq. \eqref{transf} can be written in the form
    \begin{equation}
\delta \phi _{I} =\sum_{i=0}^{m}h_{Ia}^{\mu _{i}\ldots \mu _{1}}\nabla_{\mu _{i}}\cdots
\nabla_{\mu _{1}}\varepsilon^{a}\text{ ,}\label{transf2}
\end{equation}
with $h_{Ia}^{\mu _{i}\ldots \mu _{1}}$ some functions, possibly depending on the fields $\phi _{I}$ and their derivatives, which have the transformation properties under transformations of the group of invariance dictated by the indices $I$ and $a$, and whose indices ${\mu _{i}\ldots \mu _{1}}$ transform tensorially under spacetime transformations.
    \item Noether's second theorem, Eq. \eqref{bquatro}, can be written in the form
    \begin{equation}
\sum_{i=0}^{m}\left( -1\right) ^{i}\nabla_{\mu _{i}}\cdots \nabla_{\mu
_{1}}\left( \frac{\delta L_{\rm inv}}{\delta \phi _{I}}h_{Ia}^{\mu _{i}\ldots \mu
_{1}}\right) =0 \text{ ,}\label{bcinco}
\end{equation}
    \item The non-trivial on-shell identities in Theorem \ref{teo1} become relations between covariant derivatives,
    \begin{equation}
\Psi_a=\sum_{i=0}^{m}\left( -1\right) ^{i}\nabla_{\mu _{i}}\cdots \nabla_{\mu
_{1}}\left( \frac{\delta L_{\rm inv}}{\delta \zeta_{I}}h_{Ia}^{\mu _{i}\ldots \mu
_{1}}\right) \approx 0 \text{ .}\label{bcinco2}
\end{equation}
\end{enumerate}
\end{theorem}

\begin{proof}
    For the three statements:
    \begin{enumerate}[label=\arabic*.]
    \item The covariant derivative of a geometrical object living on a fiber is a combination of its ordinary derivative plus terms proportional to the connection $A_\mu$, schematically $\nabla_\mu O=\partial_\mu O+A_\mu O$. Therefore, the $\mu_i$-th covariant derivative of $O$ is a linear combination of all the ordinary derivatives up to order $\mu_i$. Then, Eq. \eqref{transf2} is a linear combination all the ordinary derivatives up to order $m$, that is, it can be written in the form of Eq. \eqref{transf}. The transformation properties of the coefficients $h_{Ia}^{\mu _{i}\ldots \mu _{1}}$ are determined by the transformation properties of $\delta\phi_I$ and the covariance of $\nabla_{\mu _{i}}\cdots
\nabla_{\mu _{1}}\varepsilon^{a}$.
    \item Since the covariant derivative obeys the Leibniz rule, one has
    \begin{equation}
\frac{\delta L_{\rm inv}}{\delta \phi _{I}}h_{Ia}^{\mu _{i}\ldots \mu
_{1}}\nabla_{\mu _{i}}\cdots \nabla_{\mu
_{1}}\varepsilon^a=\nabla_{\mu}V^{\mu}-\nabla_{\mu _{i}}\left(\frac{\delta L_{\rm inv}}{\delta \phi _{I}}h_{Ia}^{\mu _{i}\ldots \mu
_{1}}\right)\nabla_{\mu _{i-1}}\cdots \nabla_{\mu
_{1}}\varepsilon^a\text{ ,}\label{tf}
    \end{equation}
with
\begin{equation}
    V^{\mu_i}=\frac{\delta L_{\rm inv}}{\delta \phi _{I}}h_{Ia}^{\mu _{i}\ldots \mu
_{1}}\nabla_{\mu _{i-1}}\cdots \nabla_{\mu
_{1}}\varepsilon^a\text{ .}
\end{equation}
    Because the integral $S_{\rm inv}$ is invariant, the variation $(\delta L_{\rm inv}/\delta\phi_I)\delta\phi_I$ must itself be invariant under a group transformation, as long as it acts on the gauge parameters $\varepsilon^a$ as elements of the adjoint representation of the group, in the case of gauge groups, and it must be a scalar density with weight 1 in the case of the group of diffeomorphisms, as long as the vector fields generating the transformation are transformed themselves under further group transformations. Therefore, $V^{\mu}$ is a vector density with weight 1 for spacetime transformations and a gauge invariant for gauge transformations, hence $\nabla_{\mu}V^{\mu}=\partial_{\mu}V^{\mu}$ and the first term on the right hand side of Eq. \eqref{tf} is a boundary term that can be dropped in the partial integrations as was done in the passage from Eq. \eqref{eq9} to Eq. \eqref{eq10}, leading to Eq. \eqref{bcinco} in the same manner that Eq. \eqref{bquatro} was reached.
    \item Eq. \eqref{bcinco2} follows from Eq. \eqref{bcinco} in the very same manner that Eq. \eqref{bquatro2} follows from Eq. \eqref{bquatro}.
\end{enumerate}
\end{proof}

\section{Covariant conservation laws from Noether's second theorem}\label{sec5}

The result Eq. \eqref{bcinco2} of the previous section may contain covariant derivatives of various orders. A law of covariant conservation would emerge if only the first order covariant derivative appeared in Eq. \eqref{bcinco2}.

The infinitesimal transformation of the connection under a gauge transformation is given by
\begin{equation}
\delta A^{a}_{\mu}=\partial _{\mu }\varepsilon
^{a} + g_{\rm YM}C_{abc}A_{\mu}^b\varepsilon^{c}=\nabla_{\mu }\varepsilon^{a}\text{ ,}\label{tcon0}
\end{equation}
and the infinitesimal transformation of the metric under a diffeomorphism is given by its Lie derivative
\begin{equation}
\delta g_{\mu\nu}=-{\cal L}_\varepsilon g_{\mu\nu}=-g_{\mu\chi}\nabla_{\nu }\varepsilon
^{\chi}-g_{\nu\chi}\nabla_{\mu }\varepsilon
^{\chi}\text{ .}\label{tmet}
\end{equation}
In both cases the infinitesimal variations contain only the first covariant derivative of the group parameters.

\begin{theorem}\label{teo3}
When the group of invariance is an internal gauge group or the group of diffeomorphisms in a metric theory, and the fields $\zeta_I$ in the statement of Theorem \ref{teo1} are respectively the gauge connection or the metric, the covariant conservation laws
\begin{equation}
    \nabla_\mu j^\mu_a\approx 0
\end{equation}
hold under the cases enumerated in Theorem \ref{teo1}, with
\begin{equation}
j^\mu_a=\frac{\delta L_{\rm inv}}{\delta A_\mu^a}
\end{equation}
in the case of gauge transformations, and
\begin{equation}
j^\mu_\nu=-2g_{\nu\chi}\frac{\delta L_{\rm inv}}{\delta g_{\mu\chi}}\label{tem}
\end{equation}
in the case of diffeomorphisms. Moreover, $\nabla_\mu j^\mu_a$ is not identically zero off-shell and $j^\mu_a$ is not identically zero on-shell.
\end{theorem}

\begin{proof}
Comparing Eq. \eqref{tcon0} with Eq. \eqref{transf2}, where $\phi_I$ should be replaced by $A^b_\nu$, one gets $h_{\nu a}^{b\mu _{i}\ldots \mu
_{1}}=\delta_a^b\delta_\nu^{\mu_1}$ for $i=1$ and zero otherwise. Comparing Eq. \eqref{tmet} with Eq. \eqref{transf2}, where $\phi_I$ should be replaced by $g_{\alpha\beta}$, one gets $h_{\alpha\beta\nu}^{\mu_i\ldots\mu_1}=-\delta_\alpha^{\mu_1}g_{\beta\nu}-\delta_\beta^{\mu_1}g_{\alpha\nu}$ for $i=1$ and zero otherwise. The covariant conservation laws follow directly from Eq. \eqref{bcinco2}.
\end{proof}

\section{Examples of theories with covariant conservation laws (though not necessarily with local symmetry)}\label{sec6}

In this section we give examples of the three situations mentioned in Theorem \ref{teo3} that produce covariant conservation laws.

\subsection{Background fields}

Consider an action of the form
\begin{equation}
    S=\int_{\cal V}L_{\rm inv}\left(A_\mu^a,\psi_I\right)d^Dx\text{ ,}
\end{equation}
where the $\psi_I$ are variable fields that transform according to some representation of a gauge group, and the $A_\mu^a$ are background fields (i.e., fixed functions of spacetime), such that the Lagrangian is gauge invariant under the simultaneous transformation of both the variable fields and the background fields (transforming like the connections for the gauge group). Then we can apply Theorem \ref{teo3} and we fall in case 1 of Theorem \ref{teo1}. Therefore, Noether's second theorem provides a covariant conservation law. Theories of charged matter fields on background radiation fields fall into this category of theories

The same argument applies for an action of the form
\begin{equation}
    S=\int_{\cal V}L_{\rm inv}\left(g_{\mu\nu},\psi_I\right)d^Dx\text{ ,}
\end{equation}
where the $\psi_I$ are variable tensor fields, and $g_{\mu\nu}$ is a background metric field, such that the Lagrangian is a scalar density of weight 1 under the simultaneous action of diffeomorphisms on the variable fields and the background metric field. Theories for matter tensorial fields on a fixed curved background metric fall into this category of theories.

\medskip

\noindent{\bf Remark} {\it In theses cases, with background fields, even though the theories are not themselves invariant under local transformations, there are currents which are covariantly conserved, and they are derivable from Noe\-ther's second theorem.}

\subsection{Splitting of the action}

For an action of the form
\begin{equation}
    S=S_{\rm g}[A_\mu^a]+\int_{\cal V}L_{\rm inv}\left(A_\mu^a,\psi_I\right)d^Dx\text{ ,}\label{ymm}
\end{equation}
with all fields variable, one can apply Theorem \ref{teo3} under case 2 of Theorem \ref{teo1}.
The same happens for an action of the form
\begin{equation}
    S=S_{\rm g}[g_{\mu\nu}]+\int_{\cal V}L_{\rm inv}\left(g_{\mu\nu},\psi_I\right)d^Dx\text{ ,}\label{grm}
\end{equation}
with all fields variable.

Yang-Mills theory coupled to matter and General Relativity coupled to matter are examples of theories that fall into this category, but the general forms Eq. \eqref{ymm} and Eq. \eqref{grm} accommodate a larger number of theories, such as $f(R)$ theories coupled to matter fields.

\medskip

\noindent{\bf Remark} {\it In these cases, the presence of covariantly conserved currents is independent of the form of the terms $S_{\rm g}[A_\mu^a]$ or $S_{\rm g}[g_{\mu\nu}]$. This means that even if such terms (and hence the whole theory) are not invariant under the group of transformations considered, there are conserved currents derivable from Noether's second theorem.}

\subsection{Lagrangian containing sub-invariants}

Eq. \eqref{ymm} falls into this category when $S_{\rm g}[A_\mu^a]$ is gauge invariant. But this case is more general, accounting for any dependence of the Lagrangian on sub-invariants, not necessarily in an additive manner. In gauge theories, local gauge invariants can be constructed, such as $F^{2}=F^{a}{_{\mu \nu }}F_{a}{^{\mu \nu }}$ or $F^{3}=C_{abc}F^{a}{_{\mu \nu }}F^{b\,\mu}{_{\alpha}}F^{c\,\alpha \nu}$, and they may appear in the Lagrangian in a non-additive manner.

It is worth giving an explicit example of this case to make things clear, even if not of practical physical relevance. We choose the toy model for a non-Abelian gauge theory with Lagrangian 
\begin{equation}
    L=L_{\rm inv}=f\left(F_{\mu \nu }^{a}F_{a}^{\mu \nu }\right)\nabla^\chi\phi^*\nabla_\chi\phi\text{ ,}
\end{equation}
where $f$ is some function and
\begin{equation}
\nabla_\mu\phi=\partial_\mu\phi-{\rm i}g_{\rm YM}T_aA^a_\mu\phi
\end{equation}
is the covariant derivative of a complex scalar multiplet, $T_a$ being a Lie algebra generator in the representation of the multiplet. Now we can apply Theorem \ref{teo3}, falling into category 3 of Theorem \ref{teo1}. The equations of motion are
\begin{eqnarray}
    \frac{\delta L_{\rm inv}}{\delta A^a_{\mu}}&=&4\nabla_\nu\left(f'F_a^{\mu\nu}\nabla^\chi\phi^*\nabla_\chi\phi\right)+{\rm i}g_{\rm YM}\left(\phi^*T_a\nabla^\mu\phi-\nabla^\mu\phi^*T_a\phi\right)f\approx 0\label{pu1}\\
    \frac{\delta L_{\rm inv}}{\delta \phi}&=&-\nabla^\mu\left(f\nabla_\mu\phi\right)\approx 0\text{ .}\label{pu2}
\end{eqnarray}
One has
\begin{equation}
\left(\frac{\delta L_{\rm inv}}{\delta A^a_{\mu}}\right)_{F^2={\rm const}}={\rm i}g_{\rm YM}\left(\phi^*T_a\nabla^\mu\phi-\nabla^\mu\phi^*T_a\phi\right)f\text{ ,}
\end{equation}
which is not identically zero on-shell, and
\begin{equation}
\nabla_\mu\left(\frac{\delta L_{\rm inv}}{\delta A^a_{\mu}}\right)_{F^2={\rm const}}\approx 0\text{ ,}
\end{equation}
as can be confirmed from Eqs. \eqref{pu1}-\eqref{pu2}.

\medskip

\noindent{\bf Remark} {\it In the case of Abelian gauge theories, the field intensity $F^{\mu\nu}$ is itself a local gauge invariant. Therefore, it is enough that the Lagrangian contains the combination $F^{\mu\nu}$ of the potentials for a conserved current to exist (which in the case of Abelian gauge groups is conserved in the proper sense).}

\section{Examples of theories with local symmetry (though not necessarily with covariant conservation laws)}\label{sec7}

In this section we exhibit examples of theories with local invariance with no corresponding covariantly conserved currents via Noether's second theorem.

\subsection{Metric-affine theories}

Metric-affine theories coupled to matter fields $\psi_I$ are obtained from actions invariant under diffeomorphisms of the form
\begin{equation}
S=\int_{\mathcal{V}}\left[ L_{\rm g}\left( g_{\mu \nu },\Gamma_{\mu \nu }{^\chi}\right) +L_{\rm inv}\left( g_{\mu \nu },\Gamma_{\mu \nu }{^\chi},\psi
_{I}\right) \right] \, d^D x\text{ ,}
\end{equation}
where both the metric $g_{\mu \nu }$ and the spacetime connection $\Gamma_{\mu \nu }{^\chi}$ are to be treated as variable fields of the theory to be determined by their own equations of motion.

This action is not in the form of Theorem \ref{teo3}, since the geometric kinetic term $L_{\rm g}$ depends on both the metric and the connection. Using Eq. \eqref{tmet}
for the variation of the metric under infinitesimal diffeomorphisms and \cite{kent}
\begin{equation}
\delta\Gamma_{\mu \nu }{^\chi}=-{\cal L} _{\xi }\Gamma_{\mu \nu }{^\chi}=-\nabla_\mu\nabla_\nu\xi^\chi-R_{\gamma\mu\nu}{^\chi}\xi^\gamma+\nabla_\mu\left(T_{\gamma\nu}{^\chi}\xi^\gamma\right)
\end{equation}
for the variation of the connection, where
\begin{eqnarray}
    R_{\alpha\beta\nu}{^\mu}&=&\partial_\alpha\Gamma_{\beta\nu}{^\mu}-\partial_\beta\Gamma_{\alpha\nu}{^\mu}-\Gamma_{\alpha\nu}{^\chi}\Gamma_{\beta\chi}{^\mu}+\Gamma_{\beta\nu}{^\chi}\Gamma_{\alpha\chi}{^\mu}\\
    T_{\alpha\beta}{^\mu}&=&\Gamma_{\alpha\beta}{^\mu}-\Gamma_{\beta\alpha}{^\mu}
\end{eqnarray}
are respectivelly the curvature and torsion tensors, one can use Theorem \ref{teo2} to obtain the identity
\begin{equation}
\nabla_\mu\left(2\frac{\delta L_{\rm inv}}{\delta g_{\mu\chi}}g_{\chi\nu}-\nabla_\chi\frac{\delta L_{\rm inv}}{\delta\Gamma_{\chi\mu}{^\nu}}\right)\approx R_{\nu\alpha\beta}{^\gamma}\frac{\delta L_{\rm inv}}{\delta\Gamma_{\alpha\beta}{^\gamma}}+T_{\nu\beta}{^\gamma}\nabla_\alpha\frac{\delta L_{\rm inv}}{\delta\Gamma_{\alpha\beta}{^\gamma}}\text{ ,}
\end{equation}
which is not in the form of a law of covariant conservation.

\medskip

\noindent{\bf Remark} {\it In many situations of interest (see, for example, Ref. \cite{sot} for metric and Palatini $f(R)$ theories), the matter Lagrangian does not depend on the connection. Then, $\delta L_{\rm inv}/\delta\Gamma_{\alpha\beta}{^\mu}=0$, and a covariant conservation law still occurs.}

\medskip

\noindent{\bf Remark} {\it The presence of the metric allows one to construct the Levi-Civita connection $\tilde\Gamma_{\mu \nu }{^\chi}(g)$ and to perform the change of variables
\begin{equation}
    \Gamma_{\mu \nu }{^\chi}=\tilde\Gamma_{\mu \nu }{^\chi}(g)+K_{\mu \nu }{^\chi}\text{ ,}\label{tt}
\end{equation}
with $K_{\mu \nu }{^\chi}$ the contorsion tensor (antisymmetric in its last two indices, for a metric compatible connection, or totally arbitrary, for the most general connection). In either case, the equations of motion obtained by varying the action with respect to $\Gamma_{\mu \nu }{^\chi}$ or to $K_{\mu \nu }{^\chi}$ are equivalent because the change of variables Eq. \eqref{tt} is invertible. In this setup, one would arrive at the following, equivalent, non-trivial identity, which is not a covariant conservation law
\begin{equation}
\tilde\nabla_\mu\left(2\frac{\delta L_{\rm inv}}{\delta g_{\mu\chi}}g_{\chi\nu}+K_{\nu\alpha}{^\beta}\frac{\delta L_{\rm inv}}{\delta K_{\mu\alpha}{^\beta}}+K_{\alpha\nu}{^\beta}\frac{\delta L_{\rm inv}}{\delta K_{\alpha\mu}{^\beta}}-K_{\alpha\beta}{^\mu}\frac{\delta L_{\rm inv}}{\delta K_{\alpha\beta}{^\nu}}\right)\approx \tilde\nabla_\nu K_{\alpha\beta}{^\mu}\frac{\delta L_{\rm inv}}{\delta K_{\alpha\beta}{^\mu}}\text{ ,}
\end{equation}
with $\tilde\nabla_\mu$ the covariant derivative constructed out of the Levi-Civita connection.}

\subsection{Poincar\'e gauge theory}

In Poincar\'e gauge theory, the gauge fields are the tetrad $e_\mu^i$ and the spin-connection $\omega_{\mu i}{^j}$. The theory is invariant under local Lorentz transformations, in infinitesimal form
\begin{eqnarray}
\delta e_\mu^i &=&-e_{\mu}^k\Lambda _k{^i}\label{in1}\\
\delta \omega _{\mu i}{^j} &=&\partial _{\mu }\Lambda _i{^j}-\omega _{\mu
i}{^k}\Lambda _k{^j}+\omega _{\mu j}{^k}\Lambda _k{^i}=\nabla_{\mu }\Lambda _i{^j}\label{in2} \text{ ,}
\end{eqnarray}
with $\Lambda_i{^k}\eta_{kj}$ antisymmetric ($\eta_{ij}$ being the Lorentzian metric), together with appropriate transformation laws for the matter fields, and under local translations parameterized by $\epsilon^i$, in infinitesimal form \cite{hehl2}
\begin{eqnarray}
\delta e_\mu^i &=&-\nabla_{\mu}\epsilon^i+T_{\mu\nu}{^i}e _k^\nu\epsilon^k\label{iin1}\\
\delta \omega _{\mu i}{^j} &=&R_{\mu\nu
i}{^j}e _k^\nu\epsilon^k\label{iin2} \text{ ,}
\end{eqnarray}
with $e _i^\mu$ the inverse of the tetrad, $e _i^\mu e^j_\mu=\delta_i^j$, and
\begin{eqnarray}
    R_{\mu\nu i}{^j}&=&\partial_{\mu}\omega_{\nu i}{^j}-\partial_{\nu}\omega_{\mu i}{^j}-\omega_{\mu i}{^k}\omega_{\nu k}{^j}+\omega_{\nu i}{^k}\omega_{\mu k}{^j}\\
    T_{\mu\nu}{^i}&=&\partial_{\mu}e_{\nu}^i-\partial_{\nu}e_{\mu}^i+\omega_{\mu k}{^i}e_{\nu}^k-\omega_{\nu k}{^i}e_{\mu}^k\text{ .}
\end{eqnarray}

\medskip

\noindent{\bf Remark} {\it The theory is also invariant under diffeomorphisms generated by a vector field $\varepsilon^\mu$, which can be obtained by a combination of the former transformations with $\Lambda_i{^j}=-\omega _{\mu i}{^j}\varepsilon^\mu$ and $\epsilon^i=e^i_\mu\varepsilon^\mu$. This is reminiscent of the prescription of Bessel-Hagen \cite{del,bak,jac} for the electromagnetic field.}

\medskip

The action for Poincar\'e gauge theory coupled to matter fields $\psi_I$ has the form
\begin{equation}
    S=\int_{\mathcal{V}} \left[  L_{\rm g}\left( e_\mu^i, \omega_{\mu i}{^j}\right) + L_{\rm inv}\left( e_\mu^i, \omega_{\mu i}{^j},\psi_I\right)\right] \, d^Dx \text{ ,}\label{gen}
\end{equation}
where the kinetic term for the gauge fields $L_{\rm g}$ is constructed out of the invariants $R_{\mu\nu i}{^j}$ and $T_{\mu\nu}{^i}$.

As in the previous example, the action splits into two terms, but the kinetic term does not depend only on a metric field or a gauge potential, therefore one cannot use Theorem \ref{teo3}, but Theorem \ref{teo2} still provides the identities
\begin{eqnarray}
\nabla_\mu\left(\eta_{ki}\frac{\delta L_{\rm inv}}{\delta \omega_{\mu k}{^j}}-\eta_{kj}\frac{\delta L_{\rm inv}}{\delta \omega_{\mu k}{^i}}\right)&\approx&-e_\mu^k\left(\eta_{ki}\frac{\delta L_{\rm inv}}{\delta e_\mu^j}-\eta_{kj}\frac{\delta L_{\rm inv}}{\delta e_\mu^i}\right)\label{ju1}\\
\nabla_\mu\frac{\delta L_{\rm inv}}{\delta e_\mu^i}&\approx&-e_i^\nu\left(T_{\mu\nu}{^j}\frac{\delta L_{\rm inv}}{\delta e_\mu^j}+R_{\mu\nu k}{^j}\frac{\delta L_{\rm inv}}{\delta \omega_{\mu k}{^j}}\right)\text { .}\label{ju2}
\end{eqnarray}

In Poincar\'e gauge theory there is a naturally defined metric in spacetime,
\begin{equation}
    g_{\mu\nu}=e_\mu^ie_\nu^j\eta_{ij} \text{ ,}\label{cv1}
\end{equation}
and a naturally defined connection in spacetime can be defined using the metric condition
\begin{equation}
     \nabla_\mu e_\nu^i=\partial_\mu e_\nu^i+\omega_{\mu j}{^i}e_\nu^j-\Gamma_{\mu\nu}{^\chi}e_\chi^i=0\text{ .}\label{sc}
\end{equation}
Defining the tensors
\begin{eqnarray}
    {\cal T}^{\mu\nu}&=&\eta^{ij}e^\nu_i\frac{\delta L_{\rm inv}}{\delta e_{\mu}^{j}} \text{ ,}\\
    \cal{S}^{\mu\alpha\beta}&=&\eta^{jk}e^\alpha_ie^\beta_j\frac{\delta L_{\rm inv}}{\delta \omega_{\mu i}{^k}} \text{ ,}\label{ssss}
\end{eqnarray}
which can be related to the energy-momentum and spin current of matter respectivelly \cite{hehl2,hehl}, Eqs. (\ref{ju1})-(\ref{ju2}) become
\begin{eqnarray}
    \nabla_\mu {\cal S}^{\mu\alpha\beta} &\approx& \frac{1}{2}\left( {\cal T}^{\beta\alpha}-{\cal T}^{\alpha\beta} \right)\label{vu1}\\
    \nabla_{\mu} {\cal T}^{\mu\nu} &\approx& T^{\nu\alpha\beta}{\cal T}_{\alpha\beta}+R^{\nu\alpha\beta\gamma}{\cal S}_{\alpha\beta\gamma}\text{ .}\label{vu2}
\end{eqnarray}
In this theory the energy-momentum tensor is not symmetric, but most importantly for the purposes of this article, the identities that were obtained as a result of the invariance of one term of the action under groups of local transformations are not laws of covariant conservation.

\subsection{Weyl conformal gravity}

Weyl gravity is constructed from the action
\begin{equation}
S=\int_{\cal V}\left[\sqrt{-g}C_{\alpha\beta\gamma\delta}C^{\alpha\beta\gamma\delta}+L_{\rm inv}(g_{\mu\nu},\psi_I)\right]d^4x\text{ ,}\label{conf}
\end{equation}
where
\begin{equation}
    C_{\alpha\beta\gamma\delta}=R_{\alpha\beta\gamma\delta}+\frac{1}{2}\left(R_{\alpha\delta}g_{\beta\gamma}-R_{\beta\delta}g_{\alpha\gamma}-R_{\alpha\gamma}g_{\beta\delta}+R_{\beta\gamma}g_{\alpha\delta}\right)+
    \frac{1}{6} R\left(g_{\alpha\gamma}g_{\beta\delta}-g_{\alpha\delta}g_{\beta\gamma}\right)
\end{equation}
is the Weyl tensor in $D=4$ spacetime dimensions. The first term in the action is invariant under diffeomorphisms and under conformal transformations $g_{\mu\nu}\to\exp{(\lambda)} g_{\mu\nu}$, which infinitesimally take the form
\begin{equation}
    \delta g_{\mu\nu}=\lambda g_{\mu\nu}\text{ .}\label{tcon}
\end{equation}
The matter part of the action Eq. \eqref{conf} can be made invariant under the same transformations with appropriate conformal weights for the matter fields $\psi_I$ \cite{bir}.

The action is split into two terms, one depending on the metric only and the other depending on both the metric and the matter fields, so that the second case of Theorem \ref{teo1} holds. Therefore, there are five non-trivial identities involving the energy-momentum tensor Eq. \eqref{tem}. But there are only four covariantly conserved currents $j^\mu_\nu$, with $\nabla_\mu j^\mu_\nu\approx 0$, because only the transformations produced by diffeomorphisms, Eq. \eqref{tmet}, fall into the category of Theorem \ref{teo3}. The remaining transformation, Eq. \eqref{tcon}, produces, according to Theorem \ref{teo1}, an identity which is not a conservation law, namely,
\begin{equation}
j^\mu_\mu =0\text{ .}
\end{equation}

Indeed, since there is only one parameter in Eq. \eqref{tcon}, the index $a$ in Eq. \eqref{bquatro2} is unnecessary, and since Eq. \eqref{tcon} does not involve derivatives of $\lambda$, $m=0$, and Eq. \eqref{bquatro2} becomes simply $\Psi=(\delta L_{\rm inv}/\delta \zeta_I)f_I$. In this case $\zeta_I=g_{\mu\nu}$, so that $\Psi=(\delta L_{\rm inv}/\delta g_{\mu\nu})f_{\mu\nu}$. From Eq. \eqref{tcon} we can read $f_{\mu\nu}=g_{\mu\nu}$, hence $\Psi=(\delta L_{\rm inv}/\delta g_{\mu\nu})f_I=j^{\mu\nu}g_{\mu\nu}=j_\mu^\mu\approx 0$.

\subsection{$f(R,T)$ gravity theory}

Alternatives to General Relativity have been proposed in which geometry and matter can be non-minimally coupled to each other. To illustrate it, let us consider the so-called $f(R,T)$ gravity \cite{har}, whose action takes the following form
\begin{equation}
    S=\int_{\mathcal{V}}\left[ \sqrt{-g}f\left( R,T\right)+L_{\rm m}\left(g_{\mu\nu},\psi_I\right)\right] \, d^4x \text{ ,}
\end{equation}
where the arguments of the first term are the Ricci scalar $R$ and
\begin{equation}
    T=\frac{2g_{\mu\nu}}{\sqrt{-g}}\frac{\delta L_{\rm m}}{\delta g_{\mu\nu}} \text{ .}
\end{equation}
This is a metric theory which is invariant under diffeomorphisms, where none of the cases identified in Theorem \ref{teo1} occurs (since both terms of the action may depend on all the variable fields, it does not fall into the second case), there are no background fields, and there are no sub-invariants depending on a subset of the fields. Therefore, no covariant conservation law can be derived using Noether's second theorem.

\medskip

\noindent{\bf Remark} {\it In the vast majority of cases studied in these types of theory \cite{har}, in order to reduce the theory back to General Relativity in the appropriate regimes, the choice is made for the function $f$ to have the form
\begin{equation}
    f\left( R,T\right)=f_1\left( R\right)+f_2\left( R,T\right) \text{ .} \label{funcao}
\end{equation}
In such circumstances, the Lagrangian splits into two terms, one of them, $f_1$, depending only on the metric, and the other being $L_{\rm inv}=\sqrt{-g}f_2+L_{\rm m}$. Then the conditions of Theorem \ref{teo3} are met, falling in case 2 of Theorem \ref{teo1}, and there is a law of covariant conservation,
\begin{equation}
\nabla_\mu\left[\frac{\delta\left(\sqrt{-g}f_2\right)}{\delta g_{\mu\nu}}+\frac{\delta L_{\rm m}}{\delta g_{\mu\nu}}\right]\approx 0\text{ .}\label{temu}
\end{equation}
In these theories, it is common to write $L_{\rm m} = \sqrt{-g} \mathcal{L}_{\rm m}$, with being $\mathcal{L}_{\rm m}$ the ``matter Lagrangian'' constructed with the minimal coupling prescription, and the majority of the authors give $T^{\mu\nu} =  2 (\delta \sqrt{-g}\mathcal{L}_{\rm m} / \delta g_{\mu\nu})/\sqrt{-g}$}  the physical interpretation of the energy-momentum of matter, which then is not covariantly conserved, while Eq. \eqref{temu} holds.

\section{Special cases}\label{sec8}

When the conditions established in Theorem \ref{teo3} for the existence of a covariantly conserved current are not met, it may still be possible to find such a current, because of the specific form of the equations of motion. If an equation of motion is of an algebraic nature, then it can be solved for that variable, and after plugging the solution into the original Lagrangian, the resulting effective Lagrangian may satisfy the conditions of Theorem \ref{teo3}. There may also be ``artificial'' constructions, not following from Noether's second theorem, and of questionable physical relevance, which obey covariant conservation laws.

\subsection{Einstein-Cartan theory}

The lowest order Poincar\'e gauge theory is obtained by choosing in Eq. \eqref{gen}
\begin{equation}
    L_{\rm g}=\frac{1}{2G}\eta^{ij}e\,e^\mu_ie^\nu_kR_{\mu \nu
j}{^k}\text{ ,}
\end{equation}
with $e$ the determinant of $e_\mu^i$ and $G$ the gravitational constant. The resulting theory (Einstein-Cartan theory), without coupling to matter, is equivalent to General Relativity. Coupling to a Dirac spinor field can be introduced via the Lagrangian
\begin{equation}
L_{\rm inv}=e\left[\frac{\rm i}{2}e^\mu_i\left( \bar{\psi}\gamma ^{i}\nabla_{\mu }\psi -\overline{\nabla_{\mu }\psi}\gamma ^{i}\psi \right) -M\bar\psi \psi\right]\text{ ,}
\end{equation}
where the covariant derivative of the Dirac spinor $\psi$ is given by
\begin{equation}
\nabla_{\mu }\psi=\partial _{\mu }\psi +\frac{\rm i}{8}\eta_{jk}\omega _{\mu i}{^k}\sigma ^{ij}\psi \text{ ,}
\end{equation}%
with $\sigma ^{ij}={\rm i}\left[\gamma^i,\gamma^j\right]/2$.

The action for the theory, $S=\int_{\cal V}(L_{\rm inv}+L_{\rm g})\,d^4x$, is of first order in the derivatives of the spin-connection $\omega _{\mu i}{^j}$, and linear in its first derivatives. Therefore, the equations of motion for the spin-connection are algebraic in the spin-connection,
\begin{equation}
\frac{\delta (L_{\rm inv}+L_{\rm g})}{\delta\omega_{\mu i}{^j}} =S^{\mu\alpha\beta}e_\alpha^ie_\beta^k\eta_{kj}+\frac{1}{G}\nabla_{\nu }\left[\eta^{ik}e\left( e^\mu_ke^\nu_j-e^\nu_ke^\mu_j\right)\right]\approx 0\label{ome} \text{ ,}
\end{equation}
with
\begin{equation}
    S^{\mu\alpha\beta}=\frac{1}{2}ee^\mu_ie^\alpha_je^\beta_k\eta_{lm}\epsilon^{ijkl}\bar{\psi}\gamma _{5}\gamma^m\psi \text{ ,}
\end{equation}
and they can be solved, with solution
\begin{equation}
\omega _{\mu i}{^j}\approx \tilde{\omega}_{\mu i}{^j}-\frac{G}{2e}S_{\mu\alpha}{^\beta}e^\alpha_ie_\beta^j\text{ ,}\label{sco}
\end{equation}
where $\tilde{\omega}_{\mu i}{^j}$ is the spin-connection corresponding to the Levi-Civita spacetime connection via the metric condition Eq. \eqref{sc}, given by
\begin{equation}
2\tilde\omega _{\mu i}{^j}=\eta^{jk}\eta_{mn}e_\mu^me_{i}^{\chi}e_{k}^{\nu }\left( \partial _{\chi
}e_\nu^n-\partial _{\nu }e_\chi^n\right) -e_{i}^{\nu }\left(
\partial _{\mu }e_\nu^j-\partial _{\nu }e_\mu^j\right) +\eta_{ik}\eta^{jl}e_{l}^{\nu }\left( \partial _{\mu }e_\nu^k-\partial _{\nu }e_\mu^k\right) \label{scon}\text{ ,}
\end{equation}

Eq. \eqref{sco} translates, in terms of the spacetime connection, into
\begin{equation}
\Gamma _{\mu \nu }{^\alpha }\approx \tilde{\Gamma}_{\mu \nu }{^\alpha }-\frac{G}{2e}%
S_{\mu \nu }{^\alpha }\text{ ,}
\end{equation}
with $\tilde{\Gamma}_{\mu \nu }{^\alpha }$ the Levi-Civita connection. Plugging this result into Eqs. \eqref{vu1}-\eqref{vu2}, one gets the covariant conservation law
\begin{equation}
\tilde{\nabla}_{\nu }\left[ \frac{{\cal T}^{\mu\nu }+{\cal T}^{\nu\mu }}{2}+ \frac{G}{2e}\left( {\cal S}^\mu{_{\alpha \beta }}{\cal S}^{\alpha \beta \nu }-\frac{1}{2}%
{\cal S}_{\gamma \alpha \beta }{\cal S}^{\alpha \beta \gamma }g^{\mu\nu
}\right) \right]\approx 0\text{ ,}\label{ult}
\end{equation}
where $\tilde{\nabla}_{\nu }$ is the covariant derivative with respect to
the Levi-Civita connection $\tilde{\Gamma}_{\mu \nu }{^\alpha }$.

This result should not surprise us. It is known that, when one can solve
algebraically the equations of motion for one variable, it is permissible to
replace this variable as a function of the others in the original
Lagrangian, thus eliminating the variable altogether. The resulting
action, in this case
\begin{equation}
    S_{\rm eff}=\int_{\cal V}\left(\tilde L_{\rm inv}-\frac{1}{2}%
{\cal S}_{\gamma \alpha \beta }{\cal S}^{\alpha \beta \gamma }+\tilde L_{\rm g}\right)d^4x\text{ ,}\label{gig}
\end{equation}
with the tildes meaning that $\omega _{\mu i}{^j}$ is to be replaced by $\tilde\omega _{\mu i}{^j}$, provides the same solution for the remaining variables. The only remaining gauge field is the tetrad field, and the theory becomes purely metric, hence Theorem \ref{teo3} can be applied. Indeed, since
torsion vanishes for the Levi-Civita connection, Eq. \eqref{iin1} tells us that the infinitesimal variation of the tetrad is proportional to the covariant derivative of the gauge parameter. The covariantly conserved tensor in Eq. \eqref{ult} is simply the variational derivative of the effective matter Lagrangian (the first two terms in Eq. \eqref{gig}) with respect to the tetrad.

\medskip

\noindent{\bf Remark} {\it It should be noted that the emergence of a covariantly conserved tensor was only possible due to the fact that the equation of motion for the spin-connection is algebraic in the spin-connection. In the general case, Eq. \eqref{gen}, it is not possible to solve algebraically for the spin-connection, and no such conservation law can be derived.}

\medskip

\noindent{\bf Remark} {\it It can be seen from Eq. \eqref{gig} that the effective Lagrangian contains an extra quartic term in the spinor filed, when compared with the action built directly from the spin-connection $\tilde\omega _{\mu i}{^j}$ taken as a function of tetrad. But it is more in the spirit of gauge theories to treat the spin-connection as a dynamical variable in par with the tetrad \cite{hehl2,hehl}. In practice, the effects of this change are supposed to be felt only at extremely high densities of matter, not currently achievable.}

\subsection{Non-Noetherian currents}

Even when Noether's theorem cannot produce conser\-ved currents, it is possible, in certain cases, to find expressions whose covariant divergence vanishes on-shell. Diffeomorphism invariant theories constructed out of matter tensors and the metric are one such case, with an action with the general form
\begin{equation}
S=\int_{\mathcal{V}}\sqrt{-g}F\left( g_{\mu \nu },\psi _{I}\right) \, d^Dx \text{ .}
\end{equation}
The variational derivative of the Lagrangian $L=\sqrt{-g}F$ with respect to the metric is
\begin{equation}
\frac{\delta L}{\delta g_{\mu \nu }}=\sum_{n=0}^{N}\left( -1\right)
^{n}\partial _{\alpha _{n}\dots\alpha _{1}}\left[ \sqrt{-g}\frac{\partial F}{%
\partial \left( \partial _{\alpha _{n}\dots\alpha _{1}}g_{\mu \nu }\right) }%
\right]+\frac{1}{2}\sqrt{-g}g^{\mu \nu }F \text{ .}
\end{equation}
Therefore%
\begin{equation}
X^{\mu \nu }-G^{\mu \nu }=\frac{R}{L}\frac{\delta L}{\delta g_{\mu \nu }}\approx 0\text{ ,}
\end{equation}
where $G^{\mu \nu }=R^{\mu \nu }-g^{\mu \nu }R/2$ is the Einstein tensor and
\begin{equation}
X^{\mu \nu }=\frac{R}{L}\sum_{n=0}^{N}\left( -1\right)
^{n}\partial _{\alpha _{n}\dots\alpha _{1}}\left[ \sqrt{-g}\frac{\partial F}{%
\partial \left( \partial _{\alpha _{n}\dots\alpha _{1}}g_{\mu \nu }\right) }%
\right]+R^{\mu \nu }. \label{te1}
\end{equation}
Now $\nabla _{\nu }X^{\mu \nu }\approx 0$, by virtue of Bianchi's contracted identities.

In the case when a diffeomorphism invariant theory has an explicit dependence on the Ricci scalar $R$,
\begin{equation}
S=\int_{\mathcal{V}} L\left[ R\left( g_{\mu \nu }\right) ,g_{\mu \nu },\phi _{I}\right] \, d^Dx\text{ ,}
\end{equation}
the variational derivative of the Lagrangian with respect to the metric is
\begin{equation}
\frac{\delta L}{\delta g_{\mu \nu }}=\left(\frac{\delta L}{\delta g_{\mu \nu }}\right)_{R=\text{const}}-\frac{\delta L}{\delta R}R^{\mu \nu
}+\left( g^{\mu \nu }g^{\alpha \beta }-g^{\mu \alpha }g^{\nu \beta }\right) \nabla
_{\alpha }\nabla _{\beta }\frac{\delta L}{\delta R}\text{ ,}
\end{equation}
which can be rewritten in the form
\begin{equation}
Y^{\mu \nu }-G^{\mu \nu}=\frac{\delta L}{\delta g_{\mu \nu }} \Big/ \frac{\delta L}{\delta R}\approx 0\text{ ,}
\end{equation}
with
\begin{equation}
Y^{\mu \nu }=-\frac{1}{2}Rg^{\mu\nu}+\left( \frac{\delta L}{\delta R}\right) ^{-1}\left[\left(\frac{\delta L}{\delta g_{\mu \nu }}\right)_{R=\text{const}}+\left( g^{\mu \nu }g^{\alpha\beta }-g^{\mu \alpha }g^{\nu \beta }\right)\nabla
_{\alpha }\nabla _{\beta }\frac{\delta L}{\delta R}\right]
\text{ ,}\label{te2}
\end{equation}
thus defining yet another tensor $Y^{\mu \nu }$ with vanishing divergence on-shell, $\nabla_\mu Y^{\mu\nu}\approx 0$.

However, the physical importance of tensors such as \eqref{te1} or \eqref{te2} is at least questionable. They do not arise from Noether's theorem, they are tipically non-polynomial in the derivatives of the matter fields and they may be ill-defined in the limit of a flat spacetime. Moreover, while the covariantly conserved currents that arise from Noether's second theorem provide generators for the transformations of the matter fields $\psi_I$ in the form
\begin{equation}
Q(\varepsilon^a)=\int j^0_a\varepsilon^a\, d^{D-1}x\ ,
\end{equation}
the same does not happen with the tensors in Eqs. \eqref{te1} and \eqref{te2}, which makes their physical relevance even more doubtful.

\section{Conclusions}\label{concs}

We showed that, for gauge theories and for diffeomorphism invariant metric theories, a law of covariant conservation is present in the theory, and can be derived using Noether's second theorem, if one of the following conditions occurs:
\begin{enumerate}[label=\roman*)]
    \item The gauge potentials or the metric are external fields;
    \item The Lagrangian of the theory contains two terms, one of them depending only on the gauge potentials or the metric; that is, the dependence on the matter fields is all contained in the other term.
    \item The Lagrangian of the theory contains gauge-inva\-riant combinations of the gauge potentials only.
\end{enumerate}
When one of the above conditions is satisfied for a set of fields other than the gauge potentials or the metric, or when the local transformation is not a gauge transformation or a diffeomorphism, Noether's second theorem still provides identities that hold non trivially on-shell, though not in the form of covariant conservation laws. 

In most cases of practical interest, such as Yang-Mills theories and General Relativity, the whole theory is invariant under a group of local transformations {\it and} the second of the conditions listed above is satisfied (or the first, for background fields). It is nevertheless important to establish the general setup, as we did in this article, particularly in view of the recent attempts at exploring alternatives to General Relativity, such as Refs. \cite{har,sot}, and the following clarifications were made on the relation between covariantly conserved currents and local invariance groups:
\begin{itemize}
    \item The emergence of covariantly conserved currents from Noether's second theorem does not imply that the theory is invariant under a group of local transformations of the variable fields or spacetime, examples having been given in Sec. \ref{sec6}.
    \item Invariance of a theory under a group of local transformations does not guarantee the existence of covariant conservation laws, examples having been gi\-ven in Sec. \ref{sec7}. 
\end{itemize} 

\bigskip
{\bf \large Acknowledgements}
\medskip

We thank Francisco S. N. Lobo for discussions. MASP acknowledges support from FCT research grants\break UIDB/04434/2020 (\url{https://doi.org/10.54499/UIDB/04434/2020}) and UIDP/04434/2020 (\url{https://doi.org/10.54499/UIDP/04434/2020}), from FCT project ``BEYond LAmb\-da'' with reference PTDC/FIS-AST/0054/2021 (\url{https://doi.org/10.54499/PTDC/FIS-AST/0054/2021}), and from FCT through Fellowship UI/BD/154479/2022 (\url{https://doi.org/10.54499/UI/BD/154479/2022}). MASP also acknowledges financial support from the Luso-American Development Foundation (FLAD) under an R\&D Internship Grant, from a Fulbright Grant for Research, supported by FCT, and support from the Spanish Grants PID2020-116567GB-C21, PID2023-149560NB-C21, funded by MCIN/AEI/10.13039/501100011033. TT is supported by the FCT under contracts: UIDP/00618/2020 (\url{https://doi.org/10.54499/UIDP/00618/2020}), UID/00618/2025 and 2024.03328.CERN (\url{https://doi.org/10.54499/2024.03328.CERN}).

\end{document}